# 500 W rod-type 4x4 multi-core ultrafast fiber laser

**Arno Klenke,**[1,2,*] **Albrecht Steinkopff,**[1] **Christopher Aleshire,**[1] **Cesar Jauregui,**[1] **Stefan Kuhn,**[3] **Johannes Nold,**[3] **Christian Hupel,**[3] **Sigrun Hein,**[3] **Steffen Schulze,**[3] **Nicoletta Haarlammert,**[3] **Thomas Schreiber,**[3] **Andreas Tünnermann,**[1,2,3] **and Jens Limpert**[1,2,3]

[1]*Institute of Applied Physics, Abbe Center of Photonics, Friedrich-Schiller-Universität, Albert-Einstein-Strasse 15, 07745 Jena, Germany*
[2]*Helmholtz-Institute Jena, Froebelstieg 3, 07743 Jena, Germany*
[3]*Fraunhofer Institute for Applied Optics and Precision Engineering, Albert-Einstein-Strasse 7, 07745 Jena, Germany*
*\*Corresponding author: a.klenke@hi-jena.gsi.de*

**We present a coherently-combined femtosecond fiber CPA system based on a rod-type, Ytterbium-doped, multicore fiber with 4x4 cores. A high average power of up to 500 W (after combination and compression) could be achieved at 10 MHz repetition rate with an excellent beam quality. Additionally, <500 fs pulses with up to 600 µJ of pulse energy were also realized with this setup. This architecture is intrinsically power scalable by increasing the number of cores in the fiber.**

## 1. Introduction

Fiber laser systems employed for the amplification of ultrashort pulses have seen a rapid power evolution over recent decades. Today, average powers at the kW-level and peak powers of multiple gigawatts are available [1,2]. However, increasing the performance of ultrafast laser systems even further is challenging. In the case of fiber amplifiers, the effect of transverse mode instability is the most severe average power limiting effect [3,4]. On the other hand, the combination of the self-focusing limit and the maximum achievable stretched pulse duration in the chirped-pulse-amplification (CPA) regime (restricted due to the footprint and grating size of the stretcher/compressor) hinder a further peak-power scaling. On top of this, the maximum extractable energy from a single-core fiber also provides a hard limit for performance scaling, when taking the stretched pulse duration into account.

The coherent combination of pulses, emitted from multiple spatially separated amplifiers, is considered as a promising path to overcome these limitations [5]. This concept allows for an increase of the achievable average power and peak power linearly proportional to the number of amplifiers. In particular, the actual achievable combined power value depends on the combination efficiency and, in the best case, is equal to the number of amplifier channels. Over recent years, significant progress has been made in implementing this technology in state-of-the-art laser systems, which resulted in record performance values of more than 10 kW average power [6] and multi-mJ pulse energy emission [7]. In these systems, the footprint and the component count grow linearly with the number of combined amplifiers, and the size of a single amplifier is the main determinant of the systems overall size. Thus, the scaling of such architectures by multiple orders of magnitude is a challenging task. In order to realize systems with a very large number of channels in the future, densely scalable architectures must be developed. This can be achieved by moving from single channel components to integrated multi-channel components. The integration of multiple amplification channels into a multicore fiber (MCF), with a subsequent coherent combination of the beams, has already been investigated for tiled-aperture and filled-aperture combination in low-power experiments [8, 9]. A first combination experiment with a flexible 4x4 multicore fiber and picosecond pulses has been presented with compact beam-splitters, beam-combiners and a channel count-scalable phase stabilization system [10]. Additionally, spectral beam combination with multicore fibers has also been demonstrated [11, 12].

To reach the performance level of the previously mentioned state-of-the-art laser systems with multiple discrete amplifiers, advanced high-power compatible fiber designs can be utilized. In this letter, we present an all-glass rod-type, multicore fiber with 4x4 Ytterbium-doped cores and an embedded Flourine doped layer for pump light guidance. Filled aperture coherent combination of the output beams enables a CPA system that can deliver up to 500 W compressed average power and, in a lower repetition rate configuration, up to 600 µJ energy femtosecond pulses.

## 2. Experimental setup

In the experimental setup the ultrashort pulses are generated in a frontend system comprising a femtosecond fiber oscillator, Chirped Fiber Bragg Gratings (CFBGs) for pulse stretching, multiple preamplifiers, a spectral phase shaping system and an acousto-optical modulator (AOM) to set the pulse repetition rate. This frontend emits stretched femtosecond pulses of 1 ns duration with an average power of some tens of milliwatts depending on the repetition rate. These pulses are seeded to the final preamplifier, which delivers 3 W of average power.

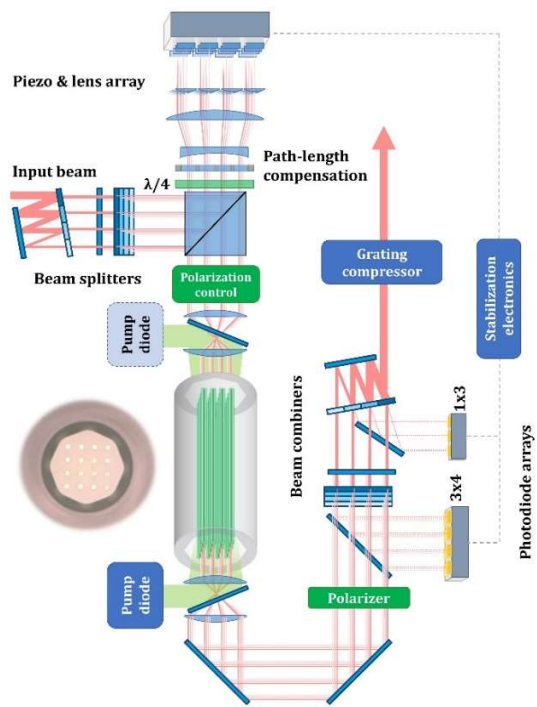

Fig. 1. Schematic setup of the rod-type, 4x4 multicore amplifier with added components for beam splitting, beam combination and phase stabilization as well as for path length adaption and polarization control. An image of the fiber end facet is shown on the left.

The output beam is then guided to the main amplifier stage as shown in Fig. 1. It is split up into a 4x4 beam array with 4 mm pitch between adjacent beams using segmented-mirror splitters (SMS) [10, 13, 14]. After that, the array is guided towards a monolithic piezo array with attached mirrors used for beam phasing in a double pass configuration by setting the quarter-wave plate (λ/4) to an angle of 45°. Due to the pitch of the piezo array being 9 mm, the incident beam array has to be magnified and, after reflection, demagnified again using a telescope. A lens array with 90 mm focal length is placed in front of the piezo array to compensate for static (which constitute the largest component) and dynamic tilts of the mirrors attached to the piezos. Additionally, glass plates of different thicknesses are inserted into the beam paths to compensate for differences in the optical path lengths between individual cores of the fiber (200 µm peak-to-peak difference); this has been optimized after installation of the fiber.

Afterwards, two waveplate arrays with small quarter-wave and half-wave plates are placed in the path of the beam array to compensate for polarization changes in the fiber. Finally, the beam array is imaged using two 4f-telescopes onto the end-facet of the rod-type 4x4 fiber, thus coupling the beam array into the fiber cores. The employed MCF was produced by using an all-glass preform with an integrated octagonal Fluorine-doped cladding (0.22 NA) that serves for pump guidance. Using deep-hole drilling of the fiber preform, the 4x4 structure for the cores is realized with a pitch-to-core-size ratio of 2.5. This is sufficient to avoid optical and thermal coupling effects between the cores [15]. After that, rods made out of Ytterbium doped glass material [16] were inserted into the stretched preform before the fiber was drawn. This manufacturing results in a cost-effective and highly scalable manufacturing process. From this preform a variety of fiber sizes were drawn by varying the drawing speed. A fiber with 21 µm core diameters (NA 0.04), 310 µm cladding diameter and 1 mm outer diameter (see Fig. 1) was chosen for this experiment. This size offers a good compromise between a sufficiently large core diameter for energy extraction and fundamental mode operation. The fiber has a length of 1.1 m, fused-silica endcaps and is mounted into a water-cooled module, which allows for efficient heat removal. The fiber is counter-pumped using a fiber-coupled pump diode emitting a maximum of 800 W at 976 nm (pump absorption 50 dB/m). For some of the experiments, an additional 600 W pump diode was used to bi-directionally pump the fiber.

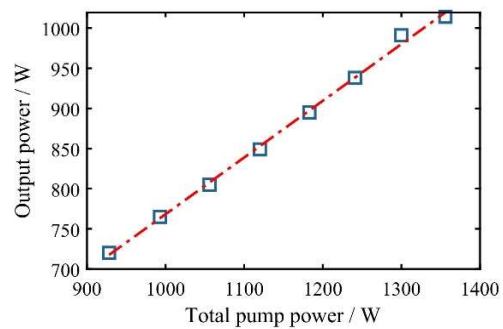

Fig. 2. Slope of the output power directly measured behind the fiber.

In the described configuration, an optical-to-optical efficiency of 70% from total pump power to multicore fiber output power is achieved at 1 kW output power measured directly at the fiber output (see Fig. 2). After amplification the emitted beam array is magnified again with two 4f-telescopes to achieve a beam array with a pitch of 4 mm (the same as on the seed side). A thin-film-polarizer (TFP) has been inserted to filter out any residual emission in the wrong polarization state. Finally, the beams are recombined using SMS elements in a configuration analogous to that of the beam-splitting stage. At this point, the non-combining parts of the beam are reflected from the SMS elements as a 3x4 array in the first combination step and as a 1x3 array in the second combination step. A fraction of these beams is guided towards two photodiode arrays with a matching number of detectors. A sequential locking scheme, as described in [10], has been employed with modulation frequencies of 6 kHz for the first row of beams and 4 kHz for the other ones. Finally, the combined beam propagates through a grating compressor with a total efficiency of around 90%. In the next sections we will describe the experimental results at two different repetition rates and discuss the observed effects in both operating regimes.

## 3. High repetition rate operation

At a high repetition rate of 10 MHz, nonlinear effects in the amplifiers can be neglected. At first, the waveplate array was optimized in the passive state (i.e. no amplification) to optimize the transmission through the TFP. With a seed power of 3 W, the average power emitted by the system was then measured at different points: behind the TFP (i.e. in front of the combination elements), after the combination stage and, finally, after compression. The combination efficiency was determined by dividing the average power after the combination stage by the average power behind the TFP. On the other hand, the compressor efficiency was calculated by the relation between the average power in front of and behind the compressor. The pump power of the counter-pumping diode was then increased up to the maximum value.

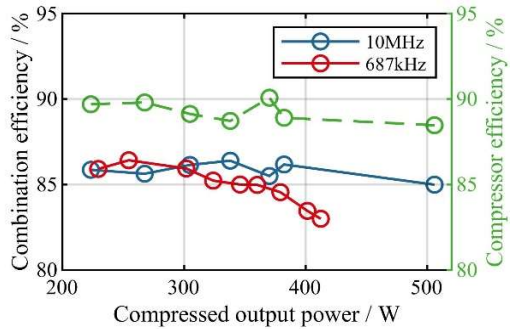

Fig. 3. Combination efficiency as a function of the compressed combined output power both for the high repetition rate (blue dotted line) and for the high pulse energy regime (red dotted line). Additionally, the compressor efficiency is shown for the first case (green dotted line).

In Fig. 3, both relations are shown as a function of the compressed combined average power for this operation regime. As can be seen, a combination efficiency of around 85% could be sustained throughout. To increase the average power even further, the co-pumping diode was also employed. However, coupling of transmitted pump light back-propagating into the diode modules, resulting in lasing at shifted wavelengths, limited the maximum pump power in the bi-directionally pumped scheme. Hence, only a fraction of the available pump power of the second diode was employed. In this configuration, 507 W of average power was achieved at a combination efficiency of ~85%. Hence, further average power scaling will be possible if laser diodes with higher power (and brightness) are employed, by choosing a fiber with a larger cladding diameter or by suppressing this lasing mechanism. In addition to the average power, the quality of both the beam and the pulses is of major importance. In Fig. 4 the $M^2$ measurement at full average power is shown.

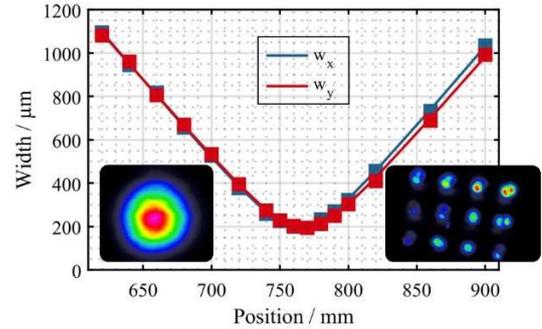

Fig. 4. $M^2$ measurement at the output of the compressor at 507 W average power and output intensity profile of the combined beam (left inset) and the non-combining emission at the first beam combiner (right inset).

A near-diffraction-limited beam quality of less than 1.1x1.1 was achieved. Measuring this value for the individual beams emitted by the fiber resulted in value of about 1.2-1.4. Hence, the coherent combination process results in an improvement of the output beam quality. This also explains the modest combination efficiency achieved, since the higher order mode content of the individual cores is filtered out, as can be seen in the right inset, thus resulting in penalty for this performance parameter. The AC duration of the combined pulse, after optimization with a phase shaper, in the frontend was 400 fs (see Fig. 5), close to the transform (TF) limit calculated from the spectrum of 375 fs.

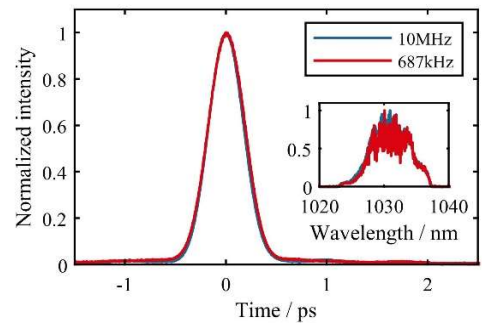

Fig. 5. Autocorrelation traces (and spectra in the inset) at the maximum output power in the high average power and high pulse energy regime.

The stability of the system was determined by measuring a photodiode trace of the system output over 20 s. A low-pass filter with a 2MHz cut-off frequency was used to filter out the pulse repetition rate while preserving high-frequency noise components. An accumulated RMS value of 0.32% was measured at full average power (see Fig. 6). A large part of this value (about 0.1%) is caused by a peak at 450 kHz, which was directly emitted from the frontend system and is not related to the main amplifier itself. Hence, further improvements should be possible in the future.

## 4. High pulse energy operation

In a second experiment, higher pulse energies were targeted by reducing the repetition rate to 687 kHz. This leads to nonlinear effects becoming more prominent which manifests itself in different observations.

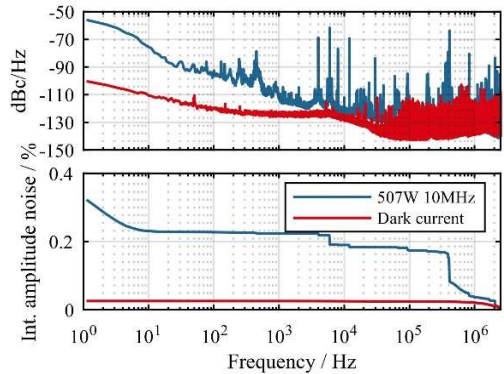

Fig. 6. Noise spectrum of the output signal behind the compressor (in blue) and dark current of the photodiode (in red). The lower plot is the integrated amplitude noise.

On the one hand, nonlinear polarization rotation was observed in the cores, reducing the TFP throughput from 92% in the high repetition rate case to 86%. This was compensated with the waveplate array by optimizing the transmission through the TFP again at the targeted average power and pulse energy regime. While this results in the same output polarization for each core, the input polarizations will be different. Therefore, the polarization states during propagation and the encountered nonlinearities are different for each core. Hence, their accumulated B-Integral is different, which results in an additional loss of combination efficiency. Multicore fibers with polarization-maintaining structures will be a way to solve this issue in the future. On the other hand, the absolute B-Integral imprints a spectral phase on each pulse, which has to be compensated by the phase shaper to keep an acceptable pulse quality. An AC duration of 420 fs (385 fs TF-limit) could be achieved after this optimization at the highest pulse energy. In Fig. 3, the combination efficiency for 687 kHz operation is shown. Please note that only single-side pumping was employed in this configuration, as the pump diode lasing threshold occurred earlier and resulted only in small improvement. This is caused by the lower amplification efficiency due to the higher pulse energy extraction, which results in more transmitted pump light. As can be seen, the combination efficiency drops slightly. Nevertheless, a value of 83% could be achieved at 412 W average power, corresponding to 600 µJ of total output pulse energy.

## Summary

In summary, we have built a femtosecond fiber CPA system based on coherently combining the output of an all-glass, 4x4 multicore fiber. This system allows for compressed output average powers of up to 500 W. Additionally, pulse energies of up to 600 µJ at 410 W average power could be achieved. To the best of our knowledge, these are the highest performance values for a multicore fiber based coherently combined femtosecond laser system to date. A maximum combination efficiency of 85% was measured, slightly dropping to 83% when operated at higher pulse energies.

The performance-scaling of this architecture can be achieved by increasing the core count and/or the core diameter of the fiber. In this respect, while the pulse energy scaling potential is clearly given by the total mode area (determined by the total number of cores and their area), the average power scaling potential is less obviously linked to the number of cores. However, a favorable average power scaling dependency on the number of cores up to the multi-kW regime has already been theoretically predicted [17]. This paves the way to the development of high-power, high-energy ultrafast fiber laser systems in the future.

**Data availability.** Data underlying the results presented in this paper are not publicly available at this time but may be obtained from the authors upon reasonable request.

**Disclosures.** The authors declare no conflicts of interest.

**Funding.** German Federal Ministry of Education and Research (funding program Photonics Research Germany, contract no. 13N15244, PINT). Fraunhofer Cluster of Excellence Advanced photon sources (CAPS). European Research Council (ERC) under the European Union's Horizon 2020 research and innovation programme (grant 835306, SALT). Free State of Thuringia and the European Social Fund with the project RATI (2018 FGR 0099). Thüringer Aufbaubank (TAB Forschergruppe TAB-FGR0074). C. Jauregui acknowledges funding by the Deutsche Forschungsgemeinschaft (DFG, German Research Foundation); 416342637. C. Aleshire also acknowledges funding by the DFG (259607349/GRK2101).